\documentclass[aps,pre,twocolumn,superscriptaddress,showpacs]{revtex4}
\usepackage{epsfig}
\usepackage{amsmath} 
\usepackage{times}

\begin{document}

[Phys. Rev. E {\bf 68}, 036105 (2003)]

\title{Large-scale structural organization of social networks}

\author{Adilson E. Motter}
\email{motter@mpipks-dresden.mpg.de}
\affiliation{Department of Mathematics,
Arizona State University, Tempe, Arizona 85287, USA}
\affiliation{Max Planck Institute for the Physics of Complex Systems, N\"othnitzer Strasse 38,
01187 Dresden, Germany}

\author{Takashi Nishikawa}
\affiliation{Department of Mathematics, 
Arizona State University, Tempe, Arizona 85287, USA}
\affiliation{Department of Mathematics, Southern Methodist
University, Dallas, Texas 75275, USA}

\author{Ying-Cheng Lai}
\affiliation{Department of Mathematics, 
Arizona State University, Tempe, Arizona 85287, USA}
\affiliation{Departments of Electrical Engineering and Physics,
Arizona State University, Tempe, Arizona 85287, USA}

\begin{abstract}
The characterization of large-scale structural organization of social
networks is an important interdisciplinary problem. We show, by using
scaling analysis and numerical computation, that the following
factors are relevant for models of social networks: 
the correlation between friendship ties among people
and the position of their social groups, as well as the correlation
between the positions of different social groups to which a person
belongs. 
\end{abstract}
\maketitle

Application of concepts and tools from physics
to the understanding of large-scale structural organization of social 
networks is an interesting interdisciplinary topic. This is 
particularly so when considering that a social network is typically 
a complex network \cite{5} that possesses the small-world property \cite{4}.
There is now a large recent literature concerning complex networks, 
for which ideas and methodologies from statistical and nonlinear physics 
have proven to be useful \cite{5,4}. The purpose of this Letter is to 
present a quantitative analysis elucidating some fundamental ingredients
required for models of complex, social networks. 

The problem that motivates our analysis is the {\it small-world
phenomenon}, according to which any two people are connected by a
short chain of acquaintances \cite{1,2,3}.  Although sociological in
origin, the small-world phenomenon has been observed in a variety of
natural and man-made systems \cite{4,5}, with examples ranging from
word association \cite{6} to the Internet \cite{7}. The {\it
existence} of short paths in these systems has been successfully
described by network models with some degree of randomness
\cite{18,9,8-9}.  However, since short paths are present in most
random networks, it is not clear which models are sociologically more
plausible, and the real structure of the network of social ties still
remains widely unknown.

A more involved and entirely different issue
concerns the {\it discovery} of short paths based only on local
information, such as in a process of target-search
\cite{10,10-11,11,11-12,12,13}, which has been only partially
understood.  In particular, the phenomenon of quick and easy
identification of acquaintances has not been explained yet at a
fundamental level. 
When two people are introduced to each other, they are naturally
inclined to look for social connections that can identify them with
the newly introduced person.  In this process, they often discover
that they share common friends, that their friends live or work in the
same place, etc.  Considering the typically large size of the
communities and the limited number of acquaintances a person has, this
happens with a surprisingly high probability, even if we accept that
people systematically underestimate the likelihood of coincidences.
The often successful identification of acquaintances is even more
striking in view of the very small number of friends usually mentioned
in an introductory conversation.
As we show, the existence of short paths connecting
people, although to some extent necessary, is not a sufficient
condition for the frequent identification of common friends to occur,
even when we consider that strangers who meet are more likely to have
mutual friends than randomly selected people. Indeed, the networks
that account for this phenomenon contain both random and {\it regular}
components and are necessarily {\it highly correlated} (to be
described below). This result constrains the possible
structure of the actual network of acquaintances and provides
insight into the properties of social networks.  These properties are
potentially relevant to a variety of other networks as well.

A class of social network models has been recently proposed by Watts,
Dodds, and Newman (WDN) \cite{11}, which can explain the letter-sending
experiment of Travers and Milgram \cite{14}. In this model, people are
organized into groups according to their social characteristics. These
groups in turn belong to groups of groups and so on, forming a {\it
hierarchy} of social structure.  A different hierarchical scheme is
defined for each social characteristic \cite{15}, which is assumed in the
WDN-model to be {\it completely independent} of one another.  The
network is then constructed using the notion of social distance
defined in terms of this set of hierarchies.  However, social groups
are often correlated. For example, people who work or study together
are more likely to engage in other activities together.  As we show, a
proper level of correlation among social groups is the key to
discovering social connections between individuals.

\begin{figure}[t]
\begin{center}
\epsfig{figure=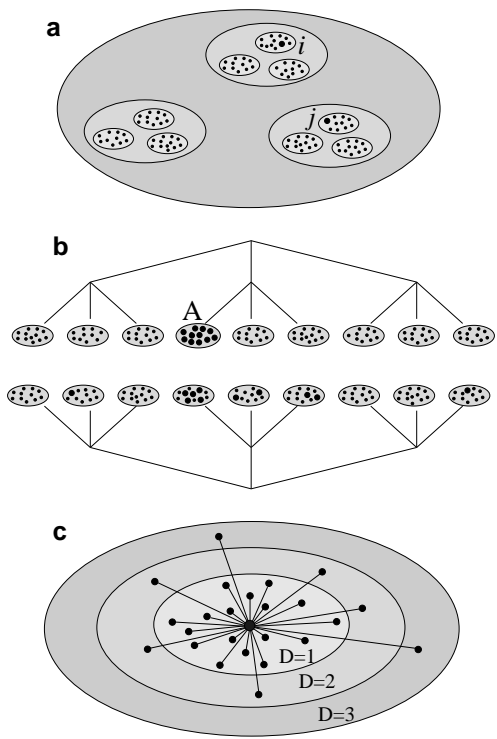,width=6.5cm}
\caption{ 
Model of social network.
(a) People (dots) belong to groups (ellipses),
which in turn belong to groups of groups and so on. 
The largest group corresponds to the entire community.
As we go down in this hierarchical organization, each group represents a set of people
with increasing social affinity.
In the  example, there are $l=3$ hierarchical levels, each representing
a subdivision in $b=3$ smaller groups, and the lowest groups are composed of $g=11$
people, on average. This defines a social hierarchy.
The distance between the highlighted individuals $i$ and $j$ in this hierarchy is 3.
(b) Each hierarchy can be represented as a tree-like structure.
Different hierarchies are correlated, in the sense that distances that are short
along one of them are more likely to be short along the others as well.
The figure shows an example with $H=2$ hierarchies, where highlighted in the second
hierarchy are those people belonging to group $A$ in the first one.
(c) Pairs of people at shorter social distances are
more likely to be linked by social ties, which can
represent either friendship or acquaintanceship ties (we do not distinguished them here
because the ones that are relevant for the problem in question may depend on the social context).
The figure shows, for a person in the network, the distribution
of acquaintances at social distance $D=1, 2,$ and $3$, where $D$ is the minimum over
the distances along all the hierarchies.
}
\label{fig1}
\end{center}
\end{figure}

{\it Network Model} -- We consider a community of $N$ people, which
represents for instance the population of a city.  People in this
community are assumed to have $H$ relevant social characteristics that
may correspond to professional or private life attributes.  Each of
these characteristics defines a nested hierarchical organization of
groups, where people are split into smaller and smaller subgroups
downwards in this nested structure (see Fig. \ref{fig1}a). Such a
hierarchy is characterized by the number $l$ of levels, the branching
ratio $b$ at each level, and the average number $g$ of people in the
lowest groups.  Realistic values of the parameter $g$
are on the order of tens or hundreds and represent the {\it average}
size of typical social groups, such as groups of classmates or
co-workers.  The set of groups to which a person belongs defines his
or her social coordinates, so that the social coordinates of person
$i$ are the positions $(x_i^1,...,x_i^H)$ that this person occupies in the
different hierarchies.  Given a hierarchy $h$, a distance  $d(x_i^h,x_j^h)$
along $h$ is defined for each pair of
people $(i,j)$ as the lowest level (counting from the bottom) at which
$i$ and $j$ are found in the same group (see Fig. \ref{fig1}a).
There is one such distance for each of the $H$ hierarchies.

To be concrete, we consider a network dominated by only two
hierarchies \cite{16} (generalization to higher dimensions is
straightforward).  The correlation between social groups is incorporated in the
position a person has in each hierarchy. The first
hierarchy is constructed by assigning people randomly to the lowest
groups. The second hierarchy is generated from the first by shuffling
the position of each person according to a given distribution, which
we assume to be exponential. Namely, each person is reassigned to a new
position at distance $y\in \{ 1,2,...l\}$ from the original position with probability
$P_{\beta}(y)=B\exp(-\beta y)$, where $B^{-1}={\small \sum}_{k=1}^l \exp({-\beta k})$,
so that the
constant $\beta$ characterizes the correlation between social groups.
For $\beta >-\ln b$, people who are close along one
hierarchy are more likely to be close along the other hierarchy as
well, as shown in Fig. \ref{fig1}b.  In the limit $\beta \gg -\ln b$,
both hierarchies become identical and the model reduces to the case
where $H=1$.  The WDN-model corresponds approximately to the
uncorrelated case where $\beta \approx -\ln b$.

While the social groups do not represent actual social ties,
the probability of having a link between two people depends on
the social distance between them \cite{11}. This can be modeled by
choosing a person $i$ and a hierarchy $h$ at random and linking this person
to another person $j$ at a distance $x=d(x_i^h,x_j^h)$ along $h$ with
probability 
$P_{\alpha}(x)=A\exp(-\alpha x)$, where $A^{-1}={\small \sum}_{k=1}^l \exp({-\alpha k})$,
and the correlation parameter $\alpha$ is a measure of social
affinity between acquaintances.  This process is repeated until the
average number of links per person is $n$, so that $n$ represents the
average number of acquaintances a person has.  The distance between
acquaintances will be the shortest for $\alpha \gg -\ln b$, and
typically much larger for $\alpha \approx -\ln b$ due to the uniform
distribution of ties.  Random networks are then produced
when $\alpha \approx -\ln b$, while regular networks are produced only
when $\alpha$ and $\beta$ are both large.  A realistic social network
is expected to fall somewhere in the {\it wide} region in between these two
extremes, as illustrated in Fig. \ref{fig1}c. In this region, the
networks exhibit properties of small-world networks \cite{18}, which
have been used to describe different kinds of social collaboration
networks {\cite{18,4,19,comment1}.

{\it Identification of Acquaintances} --  We assume that  a person knows another
person when he or she knows the social coordinates of the other.
When two people are introduced to each other, the information
they are likely to exchange first is that defining their social coordinates.
Next, they exchange information about their social connections,
by mentioning the social coordinates of their acquaintances.
Our goal here is to compute the probability that the newly introduced people find themselves
linked to each other through a short chain of friendship or acquaintanceship ties.

Our model of the process of introduction of two people starts with
each stranger informing the other his or her social coordinates.
Then, at each time step,
(1) one stranger cites the social coordinates of an
acquaintance {\it closest} to the other stranger (but not cited yet)
with respect to the minimum of the distances over all the hierarchies:
$D(i,j)=\min \! \mbox{}_h \; d(x_i^h,x_j^h)$; and
(2) the other stranger recognizes if the cited person is a
mutual acquaintance or an acquaintance within social distance $D=1$
of some of his or her acquaintances.
The two strangers then repeat (1) and (2) switching their
roles at every time step, until the identification in (2) succeeds or
they run out of acquaintances to cite.

The probability that two randomly chosen people have common
acquaintances, acquaintances at social distance 1 (i.e., in
the same lowest group), or acquaintances who know each other, decreases to very
small values as the network is made more and more regular, as shown in
Fig. \ref{fig3}a.  This happens because in a regular configuration,
most of the social ties connect people at short distances, and hence
the acquaintances of two people will overlap only if they are socially
close, which is unlikely to be the case for pairs of randomly chosen
people in the community.  For a random configuration, on the other
hand, there is a non-negligible probability of overlap for any two
people because their acquaintances are uniformly distributed over the
entire network.  One might then be tempted to think that the quick
discovery of common acquaintances is due to the randomness of the
network.  This, however, is far from being the case, as shown below.

\begin{figure}[t]
\begin{center}
\epsfig{figure=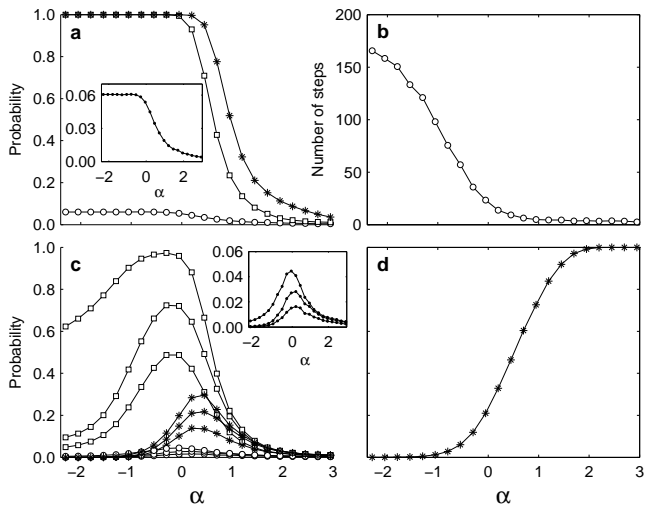, width=\linewidth}
\caption{ 
Identification of acquaintances.
(a) Probability that two randomly chosen people have
common acquaintances (circles),
acquaintances in the same lowest group (squares),
and acquaintances who know each other (stars).
Inset: blow-up of the probability of having common acquaintances.
(b) Average number of steps two strangers need to find
a common acquaintance, given that it exists.
(c) Probability that randomly chosen strangers find common acquaintances (circles),
acquaintances in the same lowest group (squares), and acquaintances in the same lowest group
who know each other (stars),
in up to $m=1,2,$ and $20$ steps (from bottom to top).
Inset:  blow-up of the probability of finding common acquaintances.
(d) Probability that two people in the same lowest group
know each other.
In the computations shown, we set $\beta=\alpha$, but similar results were
observed for any path in the $\alpha\beta$-plane interpolating from random to regular networks.
The other parameters are $N=10^6$, $n=250$, $g=100$, $b=10$, and $H=2$,
which makes $l=5$.
The size $N$ of the networks is typical for the population of a large metropolitan
city, and the average number of acquaintances $n$ is consistent with empirical values \cite{22}.
}
\label{fig3}
\end{center}
\end{figure}

In Fig. \ref{fig3}b we display the average number of steps needed for
randomly chosen strangers to find a common acquaintance, {\it given
that it exists}.  In contrast to Fig. \ref{fig3}a, the number of steps
increases sharply as the randomness of the network is made larger,
which means that it is extremely difficult to identify common
acquaintances in random networks. Indeed, while in the regular regime
only a few steps are required on average, in the random regime it
requires well over a hundred steps.  This happens because, in the
random limit, the social coordinates of a person are completely
uncorrelated with his or her social ties, and hence do not give any
clue for the position of the person's acquaintances. Accordingly,
since only a few among $n$ acquaintances are typically shared with the
other person, they need to go through many steps to identify the
overlap. When there is a single common acquaintance, the average
number of steps approaches $n$, which is on the order of hundreds.
Therefore, the probability that two people have common acquaintances
is larger for random networks, but if common acquaintances exist it is
easier for these people to find them when the underlying network is
regular.

Gathering all these together, we have that the identification of
acquaintances is most probable in between these two extremes, which is
verified in Fig. \ref{fig3}c. In this figure, we display the
probability that two randomly chosen people identify a common
acquaintance or acquaintances in the same lowest group in $m$ or less
steps.  For small $m$, these probabilities are small in the regular
and random regimes, but they are significantly larger for a class of
networks within the small-world region.  This result expresses a
trade-off between the overlaps and the clues for people to find the
overlaps based only on local information \cite{comment2}.

In addition, our model justifies a tacit assumption people make
about the structure of the social network. When the introduced people
find that they have acquaintances in the same social group, they {\it
tacitly} assume that those two acquaintances probably know each
other. This probability is much higher for regular than for random
networks, as shown in Fig. \ref{fig3}d. In fact, in a completely
regular network the probability approaches 1 as every pair of people
at social distance 1 know each other, while in the random limit it
approaches $n/(N-1)$, which is nearly zero.  In Fig. \ref{fig3}c, we
show the corresponding probability that, in the process of
introduction, the strangers identify acquaintances at social distance
1 who actually know each other (stars).  This probability also
presents a pronounced maximum in the small-world region,
consistent with the intuition that people belonging to the same group
are likely to be acquainted.

We now consider the scaling with the system size $N$.
The probability that the identification
of acquaintances happens in the first step is
$ P_1=\sum_{k=1}^{l}\sum_{k'=1}^{k}\; Q(k) \; R(k,k') \; S(k')$,
where $Q(k)$ is the probability that the strangers are at social distance $k$ from each other,
$R(k,k')$ is the probability that the acquaintance first cited (by the first stranger)
is at social distance $k'$ from the second stranger, and $S(k')$ is the probability that the
second stranger recognizes this acquaintance either for being his or her own acquaintance or for being in the
same social group of one of them. Because of the symmetry,
the probability after 2 steps is $P_2=P_1+(1-P_1)P_1$.
To be specific, consider
the case $H=1$ for $b\gg 1$, $g\gg 1$, $n<g$, and strangers randomly chosen in the community.
Then we have
$Q(k)\approx b^{k-l}$,
$R(k,k')\approx [1-b^{k'-2}/A_k]^{B_k}-[1-b^{k'-1}/A_k]^{B_k}$,
and
$S(k')=B_{k'}/(gA_{k'})$  for  common acquaintances,
$S(k')=C_{k'}/A_{k'}$     for  acquaintances in the same lowest group,
and $S(k')=nP_{\alpha}(1)C_{k'}/(gA_{k'})$ for acquaintances in the same group who know each other,
where
$A_k=b^{k-1}$,
$B_k=n P_{\alpha}(k)$,
and $C_k=A_k[1-\exp(-B_k/A_k)]$.
The asymptotic behavior of the probabilities $P_1$ and $P_2\approx 2P_1$ is roughly 
$P\sim 1/N$,
where $N=N(b)$, as shown in  Fig. \ref{fig4} for $\alpha=0$. The same scaling is observed
for any $\alpha$. Therefore, the probabilities do not scale with the diameter of the
social network, which in the small-world region increases only
logarithmically with $N$. The rationale behind this result is that the
probability of identification of common acquaintances is limited by
the probability that common acquaintances actually exist, which for
randomly chosen pairs of people decreases as $1/N$.  Incidentally,
although the probabilities in Fig. \ref{fig3}c decrease if the number
$N$ of people is increased, a sharp maximum in the intermediate region
is always observed.

\begin{figure}[t]
\begin{center}
\epsfig{figure=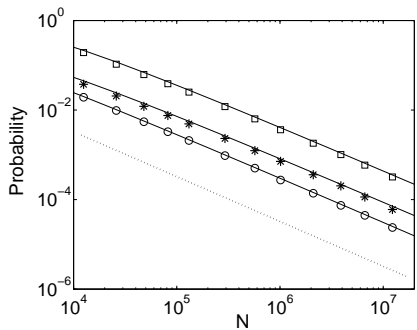,width=5.5cm}
\caption{
Probability that the identification of acquaintances happens in up to $m$ steps
as a function of the number $N$ of people in the community.
The continuous lines correspond to our theory and the symbols to the numerical verification.
We set  $m=2$, $H=1$, $n=19$, $g=20$, $l=5$, and $\alpha=0$.
The legends are the same as in Fig. \ref{fig3}c.
The dotted line is plotted for reference and corresponds to $P\sim 1/N$.
}
\label{fig4}
\end{center}
\end{figure}

{\it Conclusions} -- We have shown that the network of social ties
must be a small world {\it with} high degree of correlation for the
empirically observed frequent identification of acquaintances to be
possible.  This sheds new light on the large-scale organization of the
society, as it imposes constraints for the possible structure of
the network of acquaintances. These constraints give a criterion for
plausible models of social networks, which has implications to
issues of critical concern such as spread of diseases, homeland
defense, and propagation of influence in economic and political
systems, where the formation and behavior of social groups play
important roles. In particular, since
the dynamics of many biological agents is driven by social contacts,
reliable models of social networks are essential for efforts to reduce
the threat of biological pathogens and for making decisions in the case
of massive biological attacks. Another important conclusion of our work
is that the probability of finding a short chain of acquaintances between
two people does not scale with typical distances in the underlying network
of social ties neither with respect to system size nor across
different degrees of correlation. For instance, random networks are
usually ``smaller'' than small-world networks, and because of that
they are sometimes called themselves small-world networks.  But our
work shows that a random society would not allow people to find easily
that ``It is a small world!''

\vspace{0.1cm}

The authors thank Duncan J. Watts for illuminating conversations. 
This work was prepared at Max Planck Institute for the Physics of
Complex Systems, Dresden, and supported by AFOSR and NSF.


\begin{references}


\bibitem{5}
R. Albert and A.-L. Barab\'{a}si,
Rev. Mod. Phys. {\bf 74}, 47 (2002).

\bibitem{4}
S.H. Strogatz,
Nature (London) {\bf 410}, 268 (2001).

\bibitem{1}
I.S. Pool and  M. Kochen,
Social Networks {\bf 1}, 5 (1978).

\bibitem{2}
S. Milgram,
Psychol. Today {\bf 2}, 60 (1967).

\bibitem{3}
M. Kochen (ed.),
{\it The Small World} (Ablex, Norwood, NJ, 1989).

\bibitem{6}
A.E. Motter, A.P.S. de Moura, Y.-C. Lai, and P. Dasgupta,
Phys. Rev. E {\bf 65}, 065102 (2002).

\bibitem{7}
M. Faloutsos, P. Faloutsos, and C. Faloutsos,
Comput. Commun. Rev. {\bf 29}, 251 (1999).

\bibitem{18}
D.J. Watts and S.H. Strogatz,
Nature (London) {\bf 393}, 440 (1998).

\bibitem{9}
A.-L. Barab\'{a}si and R. Albert,
Science {\bf 286}, 509 (1999).

\bibitem{8-9}
L.A.N. Amaral, A. Scala, M. Barthelemy,  and H.E. Stanley,
Proc. Natl. Acad. Sci. USA {\bf 97}, 11149 (2000).

\bibitem{10}
J.M. Kleinberg,
Nature (London) {\bf 406}, 845 (2000).

\bibitem{10-11}
L.A. Adamic, R.M. Lukose, A.R. Puniyani,
and B.A. Huberman, Phys. Rev. E {\bf 64}, 046135 (2001).

\bibitem{11}
D.J. Watts, P.S. Dodds, and M.E.J. Newman,
Science {\bf 296}, 1302 (2002).

\bibitem{11-12}
R. Guimer\`a,  A. D\'{\i}az-Guilera, F. Vega-Redondo, A. Cabrales, and A. Arenas,
Phys. Rev. Lett. {\bf 89}, 248701 (2002)

\bibitem{12}
F. Menczer,
Proc. Natl. Acad. Sci. USA {\bf 99}, 14014 (2002).

\bibitem{13}
D.R. White and  M. Houseman,
Complexity {\bf 8}, 72 (2003).

\bibitem{14}
J. Travers and S. Milgram,
Sociometry {\bf 32}, 425 (1969).

\bibitem{15}
H.C. White,
{\it Identity and Control} (Princeton University Press, Princeton, 1992).

\bibitem{16}
H.P. Bernard, P.D. Killworth, M.J. Evans, C. McCarty, and G.A. Shelly,
Ethnology {\bf 27}, 155 (1988).

\bibitem{19}
M.E.J. Newman,
Phys. Rev. E {\bf 64}, 016131 (2001)

\bibitem{comment1}
Our model is potentially relevant to other classes of networks,
such as scientific-citation networks.
Suppose that the citation is the actual tie linking the papers.
Scientific papers can be classified according to author, subject, date, etc.,
which, along with citation, are not completely independent variables.
This defines a network with different correlated hierarchies,
similar to the social network of friends.

\bibitem{comment2}
We have focused on strangers randomly chosen from
the community, but similar results hold when the two strangers to
be introduced are correlated. In particular, if they are chosen at
social distance $z$ apart according to the distribution
$P_{\gamma}(z)\propto \exp (-\gamma z)$, where $\gamma$ is a constant,
the probabilities corresponding to Fig. \ref{fig3}c will still display
a maximum in the intermediate region, although continuously shifted to
the right as $\gamma$ is increased. Moreover, the same conclusions are
expected if the hierarchies are formed as a realization of a
stochastic branching process rather than the deterministic one
considered here.

\bibitem{22}
H.R. Bernard, E.C. Johnsen, P.D. Killworth, and S. Robinson,
Social Science Research {\bf 20}, 109 (1991).


\end{references}
\end{document}